\documentclass{article}
\usepackage[T1]{fontenc}
\usepackage[latin9]{inputenc}
\usepackage{amsmath}

\makeatletter

\newcommand{\lyxaddress}[1]{
\par {\raggedright #1
\vspace{1.4em}
\noindent\par}
}

\makeatother

\begin{document}

\title{Non-equilibrium thermodynamics as gauge fixing}


\author{So Katagiri\thanks{So.Katagiri@gmail.com}}

\maketitle


\lyxaddress{Graduate School of Arts and Sciences, The Open University of Japan,
Chiba 261-8586, Japan}
\begin{abstract}
By considering a thermodynamic force as gauge field, we extend constitutive
equations of Onsager's non-equilibrium thermodynamics to non-linear
equations. In Onsager's non-equilibrium thermodynamics, the thermodynamic
force corresponds to a pure gauge, for which the constitutive equations
are obtained by gauge fixing. If we extend the thermodynamic force
from pure gauge to physical one, we obtain the non-linear constitutive
equations of non-equilibrium thermodynamics.
\end{abstract}

\section{Introduction}

Onsager's theory is the most important one in non-equilibrium thermodynamics
with linear constitutive equations \cite{key-2,key-3}, in which constitutive
equations for currents are derived from the minimum energy dissipation
principle. Later on, this argument was supported by the path integral
representation of the probability distribution \cite{key-25,key-26,key-27,key-28,key-4}.
Onsager's theory holds in the case of linear constitutive equations,
but it is not well understood in the non-linear case (for the latest
research see \cite{key-29,key-19}). Recently, Sugamoto pointed out
with his collaborators including the present author that thermodynamic
force can be viewed as a gauge field \cite{key-1}.

In this paper we discuss this statement more definitely by means of
gauge fixing, and derive the non-linear constitutive equation by adding
the free action of the usual electromagnetism.

The paper contains the following sections. In the next section, we
review the electromagnetism in a pure gauge as an useful analogy in
the later discussions. In section 3, non-equilibrium thermodynamics
is introduced using a differential form. In section 4, we examine
gauge properties of the thermodynamic force. We extend thermodynamic
force to thermodynamical gauge field in section 5. In section 6, we
discuss this gauge theory in the path integral method. The final section
is devoted to discussion.

In addition, the paper contains the following appendices. In Appendix
A, the meaning of the time dependence of $S(a,t)$ is discussed. In
Appendix B, we examine restriction from the second law of thermodynamics.
In Appendix C, we discuss how non-linear effects work in dimensional
analysis. In Appendix D, a simple example is derived in our model.

\section{Electromagnetism}

First, we consider electromagnetism as a typical gauge theory. The
notion of a pure gauge and its gauge transformation (local phase shift)
is important in later sections in order to consider thermodynamics
as a gauge theory.

As is well known, a pure gauge $A_{\mu}$ is described as
\begin{equation}
d\theta(x)=A_{\mu}(x)dx^{\mu},
\end{equation}
where
\begin{equation}
A_{\mu}(x)\equiv\frac{\partial\theta(x)}{\partial x^{\mu}},
\end{equation}
using phase $\theta(x)$. The gauge transformation

\begin{equation}
\delta A_{\mu}(x)=\partial_{\mu}\epsilon(x)
\end{equation}
causes a phase shift
\begin{equation}
\delta\theta(x)=\epsilon(x).
\end{equation}

At the pure gauge, electromagnetic field $F$ is zero,

\begin{equation}
F=dA=dd\theta=0.
\end{equation}

For a general gauge potential $A_{\mu}$, $F$ satisfies Maxwell's
equations,

\begin{equation}
\partial_{\mu}F_{\nu\rho}+\partial_{\nu}F_{\rho\mu}+\partial_{\rho}F_{\mu\nu}=0,
\end{equation}

\begin{equation}
\partial^{\mu}F_{\mu\nu}=0.
\end{equation}

\section{Differential form of non-equilibrium thermodynamics}

A linear constitutive equation in non-equilibrium thermodynamics is
expressed as

\begin{equation}
\dot{a}^{i}=L^{ij}X_{j},\label{eq:coeq}
\end{equation}
where $a^{i}$ is extensive quantity for entropy, $X_{j}$ is thermodynamic
force for $a^{j}$, and $L^{ij}$ is Onsager's phenomenological matrix.
According to Onsager's theory \cite{key-2,key-3}, \eqref{eq:coeq}
is obtained by taking variation

\begin{equation}
\delta\left(-\frac{1}{2}R_{ij}\dot{a}^{i}\dot{a}^{j}+X_{i}\dot{a}^{i}\right)=0
\end{equation}
with respect to $\dot{a}^{i}$, where $R_{ij}$ is the inverse matrix of
$L^{ij}$,

\begin{equation}
R_{ij}L^{jk}=\delta_{i}^{k}.
\end{equation}

As its dual version, \eqref{eq:coeq} is also obtained by taking variation

\begin{equation}
\delta\left(-\frac{1}{2}L^{ij}X_{i}X_{j}+X_{i}\dot{a}^{i}\right)=0
\end{equation}
with respect to $X_{i}$.

At a first glance, thermodynamics and electromagnetism are different
theories. In thermodynamics, an important quantity, entropy $S(a,t)$,
is described in terms of extensive quantities $a^{i}$s and time $t$.
If we describe, however, entropy $S(a,t)$ in terms of differential
form, then quantities analogous to Lagrangian $L$ and gauge field
$A$ appear.\footnote{In non-equilibrium thermodynamics, the time dependence of $S(a,t)$
is natural when we consider path integral formalism and WKB approximation.
See Appendix A.}

The differential form of entropy $S(a,t)$ is expressed as

\begin{equation}
dS(a,t)=\Phi(a,t)dt+X_{i}(a,t)da^{i},\label{eq:dS}
\end{equation}
with
\begin{equation}
\Phi(a,t)\equiv\frac{\partial S(a,t)}{\partial t},
\end{equation}

\begin{equation}
X_{i}(a,t)\equiv\frac{\partial S(a,t)}{\partial a^{i}},
\end{equation}
where $\Phi(a,t)$ is called the dissipation function.

We will take the following notations:
\begin{equation}
X_{0}\equiv\Phi,
\end{equation}

\begin{equation}
a^{0}\equiv t,
\end{equation}
and $X_{\mu},~a^{\mu}$ by
\begin{equation}
\{X_{\mu}\}=X_{0},X_{1},\dots X_{N},
\end{equation}

\begin{equation}
\{a^{\mu}\}=a^{0},a^{1},\dots,a^{N}.
\end{equation}

Then, (\ref{eq:dS_xt}) is rewritten as
\begin{equation}
dS(a)=X_{\mu}(a)da^{\mu},
\end{equation}
where
\begin{equation}
X_{\mu}(a)\equiv\frac{\partial S(a)}{\partial a^{\mu}}.\label{eq:thermopuregauge}
\end{equation}

If we describe $a^{\mu}$ using parameter $\tau$, we get non-equilibrium
thermodynamical Lagrangian $L$ as follows\footnote{The parameter $\tau$ is an arbitrary parameter, and it can always be introduced into a dynamical system.}:

\begin{equation}
L(a,\frac{da}{d\tau})d\tau=X_{\mu}(a)\frac{da^{\mu}}{d\tau}d\tau.\label{eq:dStau}
\end{equation}

\section{Gauge transform property of $X_{\mu}$}

In order to know the properties of the above non-equilibrium thermodynamical
Lagrangian $L$, we have to choose an ``appropriate gauge fixing
condition'' so that we can get the Onsager's phenomenological transfer
matrix $L^{ij}$ and its constitutive equations for $X_{i}$.

The Lagrangian $L$ in (\ref{eq:dStau}) has gauge symmetry described
by the transformation

\begin{equation}
\delta X_{\mu}(a)=\partial_{\mu}\epsilon(a).\label{eq:gaugetr}
\end{equation}

The transformation (\ref{eq:gaugetr}) causes entropy shift,

\begin{equation}
\delta S(a)=\epsilon(a).\label{eq:Sshift}
\end{equation}

Since the gauge transform (\ref{eq:gaugetr}) introduces entropy's
shift (\ref{eq:Sshift}), we want to fix this gauge. To fix gauge,
we take a gauge fixing condition,

\begin{equation}
X_{0}(a)=-\frac{1}{2}L^{ij}X_{i}(a)X_{j}(a),\label{eq:gaugefix}
\end{equation}
Here, we assume $\tau=t$ to fix the parametrization invariance.

Finally, the thermodynamical Lagrangian $L$ becomes

\begin{equation}
L(a,\dot{a})dt=-\frac{1}{2}L^{ij}X_{i}(a)X_{j}(a)dt+X_{i}(a)\dot{a}^{i}dt.\label{eq:thermoL}
\end{equation}

Since $X_{i}$ is a nondynamical field, the equation of motion for
$X_{i}$ produces a constitutive equation \cite{key-2,key-3}

\begin{equation}
\dot{a}^{i}=L^{ij}X_{j}(a).\label{eq:constitutiveeq}
\end{equation}

Substituting (\ref{eq:constitutiveeq}) into (\ref{eq:thermoL}),
we obtain

\begin{equation}
L(a,\dot{a})dt=\frac{1}{2}R_{ij}\dot{a}^{i}\dot{a}^{j}dt.
\end{equation}

\footnote{This corresponds to the Rayleigh dissipation function.}

\section{Thermodynamical gauge theory}

According to gauge theory, if we consider the thermodynamic force
as a gauge potential, then the equation (\ref{eq:thermoL}), with
Onsager's phenomenological matrix, corresponds to a pure gauge theory.
On the other hand, if we introduce the physical gauge theory instead
of pure one into this thermodynamic system, then we will obtain a
different equation of motion for $X_{\mu}$, reflecting gauge-fixing
condition for the physical gauge theory. This equation of motion is
similar to Maxwell's equations in electromagnetism. Furthermore, we
can obtain the non-linear constitutive equation of non-equilibrium
thermodynamics.

So far we define thermodynamic gauge field $F$ by

\begin{equation}
F(a)=dX(a),
\end{equation}
but because $dS(a)=X(a)$, $F(a)$ is always zero,

\begin{equation}
F(a)=ddS(a)=0.
\end{equation}

Now, we will generalize this, and assume that $X_{\mu}$ is not a
pure gauge field, but a gauge field in a physical gauge theory\footnote{It's dimension is $N+1$ which is the number of components of $X$.}. The
action

\begin{equation}
S_{F}[X]=\int d^{N+1}a\,F_{\mu\nu}F^{\mu\nu}\label{eq:FF}
\end{equation}
is added to the original action with gauge fixing condition,
\begin{equation}
S_{a}[a,X,\lambda]=\int d\tau\left(X_{\mu}(a)\frac{da^{\mu}}{d\tau}-\lambda\left(X_{0}(a)+\frac{1}{2}L^{ij}X_{i}(a)X_{j}(a)\right)\right),
\end{equation}
where $\lambda$ is a Lagrange multiplier.

The equation of motion for $X_{\mu}$ is

\begin{equation}
\frac{da^{0}}{d\tau}-\lambda=\partial_{\mu}F^{\mu0}(X)=\partial_{i}E^{i}(X),\label{eq:a0equation}
\end{equation}

\begin{equation}
\frac{da^{i}}{d\tau}-\lambda L^{ij}X_{j}(a)=\partial_{\mu}F^{\mu i}(X)=\dot{E}^{i}(X)+\partial_{j}B^{ji}(X),\label{eq:aiequation}
\end{equation}
where $E_{i}$ and $B_{ij}$ are

\begin{equation}
\begin{split}E_{i}(X) & \equiv F_{i0}(X)=\partial_{i}\Phi(X)-\dot{X}_{i}\\
 & =L^{jk}X_{j}\partial_{i}X_{k}-\dot{X}_{i},
\end{split}
\end{equation}

\begin{equation}
B_{ij}(X)\equiv F_{ij}(X)=\partial_{i}X_{j}-\partial_{j}X_{i}.
\end{equation}
We call $E_{i}(X)$ and $B_{ij}(X)$ thermodynamic electric field
and thermodynamic magnetic field, respectively.

(\ref{eq:a0equation}) is solved for $\lambda$,

\begin{equation}
\lambda=\frac{da^{0}}{d\tau}-\partial_{i}E^{i},
\end{equation}
and we take static gauge $\tau=t$, then (\ref{eq:aiequation}) yields

\begin{equation}
\dot{a}^{i}=\left(1+\partial_{k}E^{k}\right)L^{ij}X_{j}+\dot{E}^{i}+\partial_{j}B^{ji}.\label{eq:dota_EB}
\end{equation}
From

\begin{equation}
\partial_{i}E^{i}=L^{il}L^{jk}\partial_{l}X_{j}\partial_{i}X_{k}+L^{il}L^{jk}X_{j}\partial_{l}\partial_{i}X_{k}-L^{ij}\partial_{i}\dot{X}_{j},
\end{equation}

\begin{equation}
\dot{E}_{i}=L^{jk}\dot{X}_{j}\partial_{i}X_{k}+L^{jk}X_{j}\partial_{i}\dot{X}_{k}-\ddot{X}_{i},
\end{equation}
we get a non-linear constitutive equation of non-equilibrium thermodynamics,
\begin{equation}
\dot{a}^{i}=L^{ij}X_{j}+\xi^{i}\label{eq:dota_Langevin}
\end{equation}
where $\xi^{i}$ is a term from the thermodynamic gauge fields:
\begin{equation}
\begin{split}\xi^{i} & =L^{ml}L^{jk}L^{in}\partial_{l}X_{j}\partial_{m}X_{k}X_{n}+L^{ml}L^{jk}L^{in}X_{j}\partial_{l}\partial_{m}X_{k}X_{n}-L^{kj}L^{il}\partial_{k}\dot{X}_{j}X_{l}\\
 & +L^{jk}L^{il}\dot{X}_{j}\partial_{l}X_{k}+L^{jk}L^{il}X_{j}\partial_{l}\dot{X}_{k}-L^{ij}\ddot{X}_{j}\\
 & +L^{jk}L^{il}\partial_{j}\partial_{k}X_{l}-L^{jk}L^{il}\partial_{j}\partial_{l}X_{k}.
\end{split}
\label{eq:dotq_A}
\end{equation}

What we have done is, by introducing the usual kinetic action of electromagnetism
into the thermodynamics, the gauge field $X_{i}(a)$ can have the
additional transverse components, in addition to the longitudinal
(or the pure gauge) component existing in the Onsager's non-equilibrium
thermodynamics. Then, the non-linear constitutive equation is obtained.

The second law of thermodynamics restricts thermodynamic gauge fields
to $\dot{E}^i+\partial_{j}B^{ji}\geq-\left(\frac{1}{2}+\partial_{k}E^{k}\right)L^{ij}X_{j}$.\footnote{See Appendix B for details. }

From the dimensional analysis in Appendix C, the non-linear term dominates
and our model with a non-linear constitutive equation becomes effective
at long relaxation time.

Thermodynamic gauge filed naturally leads to an oscillatory phenomena\footnote{As a simple example, see Appendix D.}.

\section{Path Integral}

Here, let us discuss this gauge theory in the path integral representation.
For the thermodynamic variable $a(t)$, let $P[a]$ be the probability
that a thermodynamic path will be realized with. We will call entropy
for this probability as path entropy, given by

\begin{equation}
S[a]\equiv-k_{B}\log P[a].
\end{equation}

The probability of the transition from state $(a_{i},t_{i})$ to $(a_{f},t_{f})$
can be written by path integration,

\begin{equation}
P(a_{f},t_{f}|a_{i},t_{i})=\int\mathcal{D}a\,P[a],
\end{equation}
where the path is fixed at times $t_{i}$ and $t_{f}$. The path entropy
can generally be described by the entropic Lagrangian,
\begin{equation}
S[a]=\int_{t_{i}}^{t_{f}}L(a,\dot{a})dt,
\end{equation}

\begin{equation}
L=\dot{S}=X(a)\dot{a}+\Phi(X(a),a).
\end{equation}

Here, in order to be able to regard $X$ as a free variable, we insert

\begin{equation}
\int\mathcal{D}X\,\delta\left(X(t)-\frac{\partial S}{\partial a}\right)=\int\mathcal{D}X\mathcal{D}\eta\,e^{i\int dt\,\eta(X-\frac{\partial S}{\partial a})}=1
\end{equation}
into the path integral. Then, the entropic Lagrangian is rewritten
as
\begin{equation}
L=X\dot{a}+\Phi(X,a)+i\eta\left(X-\frac{\partial S}{\partial a}\right).\label{eq:L_Xa}
\end{equation}

Solving (\ref{eq:L_Xa}) for $X$, we obtain the constitutive equation,
\begin{equation}
\dot{a}^i=-\frac{\partial\Phi}{\partial X_i}-i\eta^{i}.
\end{equation}

Here, if we take $\Phi$ as
\begin{equation}
\Phi=-\frac{1}{2}L^{ij}X_{i}X_{j},
\end{equation}
the constitutive equation yields
\begin{equation}
\dot{a}^{i}=L^{ij}X_{j}-i\eta^{i}.
\end{equation}

Substituting this expression into $\dot{S}$, we get
\begin{equation}
\dot{S}=\frac{1}{2}\dot{a}^{2}+i\eta\left(\dot{a}-\frac{\partial S}{\partial a}\right)-\frac{1}{2}\eta^{2},
\end{equation}
and solving this equation for $\eta$, we get
\begin{equation}
\dot{a}^{i}=L^{ij}\frac{\partial S}{\partial a^{j}}-i\eta^{i}.
\end{equation}

Substituting this expression into $\dot{S}$, we obtain
\begin{equation}
\dot{S}=\frac{1}{2}\left(\frac{\partial S}{\partial a}\right)^{2}-\frac{\partial S}{\partial a^{i}}i\eta^{i}.\label{eq:dotS_add_fluct}
\end{equation}

Then, the first term of $\dot{S}$ gives the the entropy increasing
law of macroscopic irreversible process, and the second term is an
effect of fluctuation.

Next, let us assume that $\Phi$ is gauge-fixed like
\begin{equation}
dS=X(a)da+\Phi(a)dt+B(\Phi(a)-\Phi(X(a),a))dt.
\end{equation}

Then, we understand that before gauge fixing, $dS$ is
\begin{equation}
dS=X(a)da+\Phi(a)dt.
\end{equation}

This action before gauge fixing is invariant under gauge transformation,
\begin{equation}
X(a)\to X(a)+\frac{\partial\epsilon}{\partial a},
\end{equation}

\begin{equation}
\Phi(a)\to\Phi(a)+\frac{\partial\epsilon}{\partial t}.
\end{equation}

Then, this theory is originally a gauge theory of
\begin{equation}
A_{\mu}=(\Phi,X).
\end{equation}

From the above discussion, the differential form representation of
non-equilibrium thermodynamics is natural from the viewpoint of the
path integral representation, and the gauge fixing is naturally introduced
to obtain a specific thermodynamical physics.

\section{Discussion}

We have assumed that the thermodynamic force is a gauge potential
and generalized this potential from a pure gauge to a physical gauge
potential. By gauge fixing (\ref{eq:gaugefix}) and adding a thermodynamical
gauge field action (\ref{eq:FF}), we have obtained a non-linear constitutive
equation for non-equilibrium thermodynamics (\ref{eq:dotq_A}). If
we choose another gauge fixing condition, we may get another thermodynamical
physics. This will be discussed elsewhere.

The viewpoint that the existing theory is some kind of gauge theory
with gauge fixing may be useful to generalize the theory. One of the
most interesting applications of this method is application to classical
mechanics, where a momentum may be regarded as a kind of statistic
\cite{key-34}.

The entropy of a thermodynamic path, which is essential for the differential
form representation of non-equilibrium thermodynamics, is also a key
to the detailed fluctuation theorem \cite{key-35}. Then, the gauge
fixing method and the gauge principle may help understand the second
law of thermodynamics.

All arguments so far are in the realm of classical theory. Reconsideration
of our discussion in the context of quantum theory is important. In
the latter case, the quantum fluctiation may be added to \eqref{eq:dotS_add_fluct}.

For other applications, application to phase transition phenomena
can be considered. In time-dependent Ginzburg-Landau theory \cite{key-36},
the free energy is expressed in terms of a complex order parameter
field and have the gauge symmetry of this parameter's phase shift.
By applying our gauge fixing method to time-dependent Ginzburg-Landau
theory, we may obtain understanding between these two gauge symmetries.

Finally, optics and non-equilibrium thermodynamics may have a closer
relationship than we have thought of, such as the invariance of the
line element of light under the gauge transformation. We will investigate
such relationship in future work.

\section*{Acknowledgments}

I am deeply grateful to Akio Sugamoto for discussions which clarify
important points on thermodynamic forces. I am indebted to Ken Yokoyama
and Shiro Komata for reading this paper and giving useful comments.
I would also like to thank Jun Katagiri, Naoki Fujimoto, Shimpei Sugita,
Noriaki Aibara and Tsukasa Yumibayashi for reading this paper.

\section*{A. The time dependence of $S(a,t)$}

In this Appendix, the meaning of the time dependence of $S(a,t)$
is discussed. As is well-known in classical mechanics \cite{key-37},
for the action
\begin{equation}
S[x]=\int dtL(x,\dot{x}),
\end{equation}

the differential of the action is equal to

\begin{equation}
dS(x,t)=pdx-Hdt.\label{eq:dS_xt}
\end{equation}

Then, we obtain
\begin{equation}
p=\frac{\partial S(x,t)}{\partial x},
\end{equation}

\begin{equation}
-H=\frac{\partial S(x,t)}{\partial t}.
\end{equation}

The last equation is Hamilton-Jacobi equation. The application of
the Hamilton-Jacobi-Equation to fluctuation phenomena is studied by
Kitahara \cite{key-38}.

If $H$ takes a constant value $E$, then we have

\begin{equation}
S(x,t)=-Et+\bar{S}(x),\label{eq:SSeparateVar}
\end{equation}

and (\ref{eq:dS_xt}) becomes

\begin{equation}
dS(x,t)=\bar{p}dx-Edt
\end{equation}

\begin{equation}
\bar{p}\equiv\frac{\partial\bar{S}(x)}{\partial x}
\end{equation}

The entropy that does not depend on time is this $\bar{S}(x)$.

\section*{B. Restriction from the second law of thermodynamics}

We want to discuss the restriction from the second law of thermodynamics
on our model with a non-linear constitutive equation.

The entropy production is

\begin{equation}
\dot{S}=\frac{\partial S}{\partial a^{i}}\dot{a}^{i}+\Phi=\frac{1}{2}L^{ij}X_{i}X_{j}+X_{i}\xi^{i}
\end{equation}

\begin{equation}
\begin{split}X_{i}\xi^{i} & =X^{2}\left(\partial_{l}X_{k}\right)^{2}+X^{2}X^{k}\partial^{2}X_{k}-X^{2}\partial\cdot\dot{X}+X^{i}\dot{X}^{j}\partial_{i}X_{j}+X^{i}X^{j}\partial_{i}\dot{X}_{j}-X^{i}\ddot{X}_{i}\end{split}
\label{eq:dotq_A-1}
\end{equation}

Since \eqref{eq:dS_xt} can be written as

\begin{equation}
S(a,t)=\Phi t+\bar{S}(a),\label{eq:SSeparateVar-1}
\end{equation}

we get

\begin{equation}
dS(a,t)=\bar{X}_{i}(a)da^{i}+\Phi dt,
\end{equation}

\begin{equation}
\bar{X}_{i}(a)\equiv\frac{\partial\bar{S}(a)}{\partial a^{i}}.
\end{equation}

Then, under (\ref{eq:SSeparateVar-1}), (\ref{eq:dotq_A-1}) becomes

\begin{equation}
\begin{split}\bar{X}_{i}\xi^{i} & =\bar{X}^{2}\left(\partial_{l}\bar{X}_{k}\right)^{2}+\bar{X}^{2}\bar{X}^{i}\partial^{2}\bar{X}_{i}\end{split}
\label{eq:dotq_A-1-1}
\end{equation}

Here we note, $\dot{E}_{i}(\bar{X})=B_{ij}(\bar{X})=0$.

If it is assumed that the second law of thermodynamics can not be
broken,

this requires the condition of

\begin{equation}
\frac{1}{2}+\left(\partial_{l}\bar{X}_{k}\right)^{2}+\bar{X}^{i}\partial^{2}\bar{X}_{i}\geq0.
\end{equation}

If we use Ruppiner metric \cite{key-39};
\begin{equation}
g_{ij}^{R}\equiv\frac{\partial^{2}\bar{S}}{\partial a^{i}\partial a^{j}}=\frac{\partial\bar{X}_{i}}{\partial a^{j}},
\end{equation}

this condition gives
\begin{equation}
2\partial^{k}\left(\bar{X}^{i}g_{ik}^{R}\right)\geq-1
\end{equation}

if $g_{ij}^{R}$ is constant, then
\begin{equation}
2g^{Rik}g_{ik}^{R}\geq-1.
\end{equation}

Because $L^{ij}$ is positive definite,
\begin{equation}
g^{Rik}g_{ik}^{R}\geq0
\end{equation}

always holds.

More generally, in term of thermodynamic electric field and thermodynamic
magnetic field, the non-liner constitutive equation is

\begin{equation}
\dot{a}^{i}=\left(1+\partial_{k}E^{k}\right)L^{ij}X_{j}+\dot{E}^{i}+\partial_{j}B^{ji}.
\end{equation}

Then, the entropy production can be given by

\begin{equation}
\dot{S}=(\frac{1}{2}+\partial_{k}E^{k})L^{ij}X_{i}X_{j}+X_{i}(\dot{E}^{i}+\partial_{j}B^{ji}).
\end{equation}

Therefore, the restriction from the second law of thermodynamics on
our model with a non-linear constitutive equation yields

\begin{equation}
\dot{E}^i+\partial_{j}B^{ji}\geq-\left(\frac{1}{2}+\partial_{k}E^{k}\right)L^{ij}X_{j}.
\end{equation}

\section*{C. Dimensional analysis}

To know physics, the dimensional analysis is important. From this
analysis, we obtain relations,

\begin{equation}
[S]=[k_{B}],
\end{equation}

\begin{equation}
[\Phi]=\frac{[S]}{[t]}=\frac{[k_{B}]}{[t]},
\end{equation}

\begin{equation}
[\frac{\partial}{\partial a^{i}}]=\frac{1}{[a^{i}]},
\end{equation}

\begin{equation}
[X_{i}]=\frac{[S]}{[a^{i}]}=\frac{[k_{B}]}{[a^{i}]},
\end{equation}

\begin{equation}
\text{\ensuremath{\frac{[a^{i}]}{[t]}}}=[L^{ij}][X_{j}]=[L^{ij}]\frac{[k_{B}]}{[a^{j}]},
\end{equation}

\begin{equation}
[L^{ij}]=\frac{[a^{i}][a^{j}]}{[t][k_{B}]},
\end{equation}

\begin{equation}
[R_{ij}]=\frac{[t][k_{B}]}{[a^{i}][a^{j}]},
\end{equation}

\begin{equation}
[F_{\mu\nu}]=\frac{[k_{B}]}{[a^{\mu}][a^{\nu}]},
\end{equation}

\begin{equation}
[F^{\mu\nu}]=\frac{[a^{\mu}]}{[k_{B}]}\frac{[a^{\nu}]}{[t^{2}]},
\end{equation}

\begin{equation}
[E_{i}]=[F_{i0}]=\frac{[k_{B}]}{[a^{i}][t]},
\end{equation}

\begin{equation}
[E^{i}]=\text{\ensuremath{\frac{[a^{i}]}{[k_{B}][t]}}},
\end{equation}

\begin{equation}
[B_{ij}]=\frac{[k_{B}]}{[a^{i}][a^{j}]},
\end{equation}

\begin{equation}
[\xi^{i}]=[\dot{a}]=\frac{[a]}{[t]}.
\end{equation}

Under these relations, we insert a basic constant of time, $\tilde{t}$
and we get

\begin{equation}
\lambda=\frac{da^{0}}{d\tau}-k_{B}\tilde{t}\partial_{i}E^{i}
\end{equation}

\begin{equation}
\dot{a}^i=(1+k_{B}\tilde{t}\partial_{k}E^{k})L^{ij}X_{j}+k_{B}\tilde{t}\left(\dot{E}^{i}+\partial_{j}B^{ji}\right)
\end{equation}

Then, if $\tilde{t}$ is long, then the non-linear term dominates
and our model with a non-linear constitutive equation becomes effective.

\section*{D. Simple example}

As a simple example, we assume,

\begin{equation}
X_{i}=g_{ij}^{R}a^{j}.
\end{equation}

\begin{equation}
\partial_{j}X_{i}=g_{ij}^{R}
\end{equation}

\begin{equation}
\begin{array}{cc}
\xi^{i}= & k_{B}\tilde{t}\left(\left(g_{mk}^{R}\right)^{2}L^{in}g_{np}^{R}a^{p}+L^{jk}L^{il}g_{jp}^{R}\dot{a}^{p}g_{lk}^{R}-L^{ij}g_{jp}^{R}\ddot{a}^{p}\right)\end{array}
\end{equation}

In this case, we can derive the following we get second order linear
differential equation,

\begin{equation}
A_{p}^{i}\ddot{a}^{p}+B_{p}^{i}\dot{a}^{p}+C_{p}^{i}a^{p}=0,\label{eq:ABC}
\end{equation}

\begin{equation}
A_{p}^{i}\equiv k_{B}\tilde{t}L^{ij}g_{jp}^{R},
\end{equation}

\begin{equation}
B_{p}^{i}\equiv\left(1-k_{B}\tilde{t}L^{jk}L^{il}g_{jp}^{R}g_{lk}^{R}\right),
\end{equation}

\begin{equation}
C_{p}^{i}\equiv-\left(1+k_{B}\tilde{t}(g_{ml}^{R})^{2}\right)L^{ij}g_{jp}^{R}.
\end{equation}

If we modify the coefficient of the first term in \eqref{eq:ABC}
to be 1, we get

\begin{equation}
\ddot{a}^{i}+D_{p}^{i}\dot{a}^{p}+E_{p}^{i}a^{p}=0,
\end{equation}

\begin{equation}
D_{p}^{i}\equiv\left(A^{-1}\right)_{j}^{i}B_{p}^{j}=(\frac{1}{k_{B}\tilde{t}}g^{Rik}R_{kp}-L^{ik}g_{kp}^{R}),
\end{equation}

\begin{equation}
E_{p}^{i}\equiv(A^{-1})_{j}^{i}C_{p}^{j}=\epsilon\delta_{p}^{i},
\end{equation}

\begin{equation}
\epsilon\equiv-\left(\frac{1}{k_{B}\tilde{t}}+(g_{ml}^{R})^{2}\right).
\end{equation}

Here we note

\begin{equation}
[D,E]=0.
\end{equation}

If $\tilde{t}\to0$, we get well known linear constitutive equation.

We take a linear transformation of $b$,

\begin{equation}
a^{p}=P_{o}^{p}b^{o},
\end{equation}

\begin{equation}
\ddot{b}^{i}+\left(P^{-1}DP\right)_{p}^{i}\dot{b}^{p}+\epsilon b^{p}=0.
\end{equation}

If we assume it can be diagonalized,

\begin{equation}
\left(P^{-1}DP\right)_{p}^{i}=\lambda_{(i)}\delta_{p}^{i},
\end{equation}

we get

\begin{equation}
\ddot{b}^{i}+\lambda_{(i)}\dot{b}^{i}+\epsilon b^{i}=0,
\end{equation}

\begin{equation}
\omega^{2}+\lambda_{(i)}\omega+\epsilon=0,
\end{equation}

\begin{equation}
\omega=\frac{-\lambda_{(i)}\pm\sqrt{\lambda_{(i)}^{2}-4\epsilon}}{2}.
\end{equation}

Under $\tilde{t}\to\infty$,

\begin{equation}
D_{p}^{i}=-L^{ik}g_{kp}^{R},
\end{equation}

\begin{equation}
\epsilon=-\left(g_{lm}^{R}\right)^{2}.
\end{equation}

If we assume for simplicity,
\begin{equation}
L^{ik}g_{kp}^{R}=\alpha\delta_{p}^{i},
\end{equation}

then

\begin{equation}
\alpha=\frac{L^{ik}g_{ki}^{R}}{N},
\end{equation}

\begin{equation}
\epsilon=-L^{lk}L^{ms}g_{lm}^{R}g_{ks}^{R}=-N\alpha^{2},
\end{equation}

\begin{equation}
P_{j}^{i}=\delta_{j}^{i}.
\end{equation}

we get
\begin{equation}
\ddot{a}^{i}-\alpha\dot{a}^{i}+N\alpha^{2}a^{i}=0,
\end{equation}

\begin{equation}
\omega=\frac{\alpha\pm\sqrt{\alpha^{2}-4N\alpha^{2}}}{2}=\frac{1\pm i\sqrt{(4N-1)}}{2}\alpha.
\end{equation}

Then, we get solution that decays while oscillating,

\begin{equation}
a(t)=a(0)e^{-\frac{\sqrt{4(N-1)}}{2}\left(\frac{L^{ik}g_{ki}^{R}}{N}\right)t}\cos\frac{1}{2}\left(\frac{L^{ik}g_{ki}^{R}}{N}\right)t\label{eq:aoci}
\end{equation}
If we take limit $N\to\infty$, the oscillating of (\ref{eq:aoci})
disappears,

\begin{equation}
a(t)=a(0)e^{-L^{ik}g_{ki}^{R}t}.
\end{equation}

Non-equilibrium macroscopic oscillatory phenomena is discussed in
the context of chemical reaction \cite{key-40}.
\end{document}